\documentclass[twocolumn]{aastex631}

\usepackage{amsmath}
\usepackage{xcolor, soul}
\usepackage{ulem}

\hypersetup{linkcolor=cyan,citecolor=teal,filecolor=cyan,urlcolor=blue}

\shorttitle{JWST View of a Rapidly Growing Dust-Obscured Quasar at $z \sim 4$}
\shortauthors{Kim \& Im}

\begin{document}

\title{JWST Spectroscopic View of a Rapidly Growing Dust-obscured Quasar at $z \sim 4$:\\ 
Effect of Dust Extinction Correction on Black Hole Mass and Eddington Ratio Estimation}

\author[0000-0002-6925-4821]{Dohyeong Kim}
\affiliation{Department of Earth Sciences, Pusan National University, Busan 46241, Republic of Korea}

\author[0000-0002-8537-6714]{Myungshin Im}
\affiliation{Astronomy Program, Dept. of Physics \& Astronomy, Seoul National University, Seoul 08826, Republic of Korea}
\affiliation{SNU Astronomy Research Center (SNU ARC), Astronomy Program, Dept. of Physics \& Astronomy, Seoul National University, Seoul 08826, Republic of Korea}

\correspondingauthor{Dohyeong Kim, Myungshin Im}
\email{dh.dr2kim@gmail.com, myungshin.im@gmail.com}

\begin{abstract}
 In the merger-driven galaxy evolution scenario, 
the central supermassive black holes (SMBHs) in dust obscured galaxies grow rapidly. 
Interestingly, a recent work \citep{suh24} on a dust-obscured galaxy, LID-568 at $z=3.965$, has shown that 
its SMBH is growing extremely fast at about 40 times of the Eddington-limited accretion rate (i.e., super-Eddington accretion). 
However, the heavy dust extinction of the host galaxy could affect the result if not corrected properly.
Here, we analyze James Webb Space Telescope (JWST) NIRSpec/IFU and MIRI spectra of LID-568.
By measuring its bolometric luminosity ($L_{\rm bol}$) and BH mass ($M_{\rm BH}$)
using an extinction-free estimator based on mid-infrared spectra, we obtain
$L_{\rm bol} = 10^{46.83\pm0.07}\,{\rm erg\,s^{-1}}$ and $M_{\rm BH} = 10^{8.43\pm0.15}\,M_{\odot}$.
The measured Eddington ratio ($\lambda_{\rm Edd}$) is 1.97$\pm$0.88,
consistent with the accretion rate at the Eddington limit;
in other words, not in super-Eddington in a significant manner.
This result underscores challenges and the importance of carefully considering dust extinction
when analyzing the BH growth in dust-obscured quasars.

\end{abstract}

\section{Introduction} \label{sec:intro}

SMBHs have been found at the centers of spheroidal galaxies,
with their masses tightly correlating with several properties of their host galaxies 
(e.g., bulge mass, stellar velocity dispersion, stellar mass, etc.; \citealt{ferrarese00,gebhardt00,kormendy13}).
It is uncertain how SMBHs were born and grew,
but several simulation studies (e.g., \citealt{hopkins06,hopkins08}) suggest that
dust-obscured quasars are a key to understand the SMBH growth process.
According to such studies, dust-obscured quasars represent the intermediate stage between
the merger-driven star-forming galaxies and unobscured quasars.
In this merger-driven galaxy evolution scenario,
the SMBHs in dust-obscured quasars are expected to be growing very rapidly,
sometimes in well-above the Eddington-limited accretion rate (super-Eddington accretion; e.g., \citealt{hopkins08}).
Several observational studies (e.g., \citealt{urrutia12,kim15,kim24b}) find that dust-obscured quasars are
accreting matters at near the Eddington limit, supporting this scenario.
Moreover, further support for the merger-driven galaxy evolution scenario comes from
several pieces of observational evidence.
For example, dust-obscured quasars exhibit 
(i) significant merging signatures in their host galaxies \citep{urrutia08,glikman15};
(ii) dusty red colors \citep{kim18a};
(iii) merging SMBH candidates \citep{kim20};
and (iv) enhanced star-forming activities \citep{georgakakis09}.
A few studies have found high redshift dust-obscured quasars with the Eddington ratios of a few \citep{tsai18,finnerty20,jun20,kim24a}. 
However, large scatters in scaling relations and uncertainties in dust-extinction correction make it difficult
to conclusively claim that such quasars are undergoing super-Eddington accretion. 
Much stronger evidence for super-Eddington accretion is therefore required.

Recently, \cite{suh24} reported the discovery of a long-sought dust-obscured quasar at $z \simeq 4$ 
that is undergoing a strong super-Eddington accretion. 
They analyzed X-ray, infrared (IR), and JWST spectroscopic data to find  the Eddington ratio of about 40 for this quasar, LID-568, 
which is well above the Eddington ratio of one for Eddington-limited accretion.

One significant difficulty in such a study is the heavy extinction 
which complicates the measurement of the black hole mass, bolometric luminosity, and Eddington ratio. 
\cite{suh24} found that the bolometric luminosity from the 5100\,$\rm \AA{}$ continuum luminosity is about a factor of 10 lower than the X-ray based value. 
They also found an Eddington ratio to be 4.4 when using an internally consistent method of deriving the Eddington ratio from H$\alpha$ line.

In this study, we measure the $L_{\rm bol}$ and $M_{\rm BH}$ values of LID-568 
using estimators designed to minimize the effects of dust extinction 
by using mid-infrared (MIR) continuum luminosity and the velocity width of the H$\alpha$ line \citep{kim23}.
Subsequently, we compare the newly-derived $\lambda_{\rm Edd}$ to other quasars at similar redshifts.
Throughout, we use a standard $\Lambda$CDM model of 
$H_{0}=70\,{\rm km\,s^{-1}}$\,Mpc$^{-1}$, $\Omega_{m}=0.3$, and $\Omega_{\Lambda}=0.7$,
which has been supported by several observational studies (e.g. \citealt{im97,planck16}).

\section{LID-568} \label{sec:LID-568}

 LID-568 is an dust-obscured quasar, 
originally identified as a near-infrared (NIR) dropout object and later as an X-ray source
in the $\it Chandra$-COSMOS Legacy Survey \citep{civano16,marchesi16}.
Follow-up JWST NIRSpec and MIRI observations revealed that its redshift is 3.965, and it has a very red continuum \citep{suh24}.
Furthermore, strong and broad hydrogen lines were detected, as shown in Figure \ref{fig:SED},
providing a supporting evidence that LID-568 is a dust-obscured quasar.
The $L_{\rm bol}$ from X-ray and $M_{\rm BH}$ from the JWST rest-frame optical spectrum provided
the $\lambda_{\rm Edd}$ of $\sim$40 \citep{suh24}.

\section{Data analysis} \label{sec:analysis}
\subsection{Spectral energy distribution fitting} \label{sec:SED}
\cite{kim23} established new $L_{\rm bol}$ and $M_{\rm BH}$ estimators for dust-obscured quasars,
which were derived based on the continuum luminosities at 3.4\,$\mu$m and 4.6\,$\mu$m 
(hereafter, $L_{\rm 3.4}$ and $L_{\rm 4.6}$, respectively) in the rest-frame.
Here, we perform an SED fitting using the JWST NIRSpec and MIRI spectra for obtaining the $L_{\rm 3.4}$.
The JWST spectra [$f(\lambda)$] are fit with a reddened quasar spectrum [$Q(\lambda)$],
as $f(\lambda) = CQ(\lambda)$, where $C$ is the normalization constant.
Here, the $Q(\lambda)$ is reddened from an unobscured quasar spectrum [$Q_{\rm 0} (\lambda)$; \citealt{assef10}]
using a reddening law based on Small Magellanic Cloud bar dust extinction curve with $R_V = 2.74$ \citep{gordon03}.
Note that even if the Galactic dust extinction curve with $R_V = 3.1$ \citep{fitzpatrick99} is adopted instead,
the derived $L_{\rm 3.4}$ shows no significant ($<5\,\%$) difference.

Meanwhile, \cite{kim23} used several types of galaxy spectra in their SED fitting, but for LID-568, 
we adopt the result where only the $Q(\lambda)$ component is employed.
We attempted the SED fitting including the galaxy spectra as done by \cite{kim23},
but the galaxy contributions to the SED are found to be negligible.
Moreover, we adopt the quasar spectrum of \cite{assef10} for this work,
as it yields the best reduced chi-square fit compared to using other quasar spectra from \cite{richards06} and \cite{krawczyk13}.
However, regardless of which quasar spectrum is used for the fitting, 
the measured $L_{\rm 3.4}$ does not vary significantly \citep{kim23},
and the discrepancy is added as an uncertainty in quadrature to
its original uncertainties obtained from the SED fit.

\begin{figure*}
	\centering
	\figurenum{1}
	\includegraphics[scale=0.3]{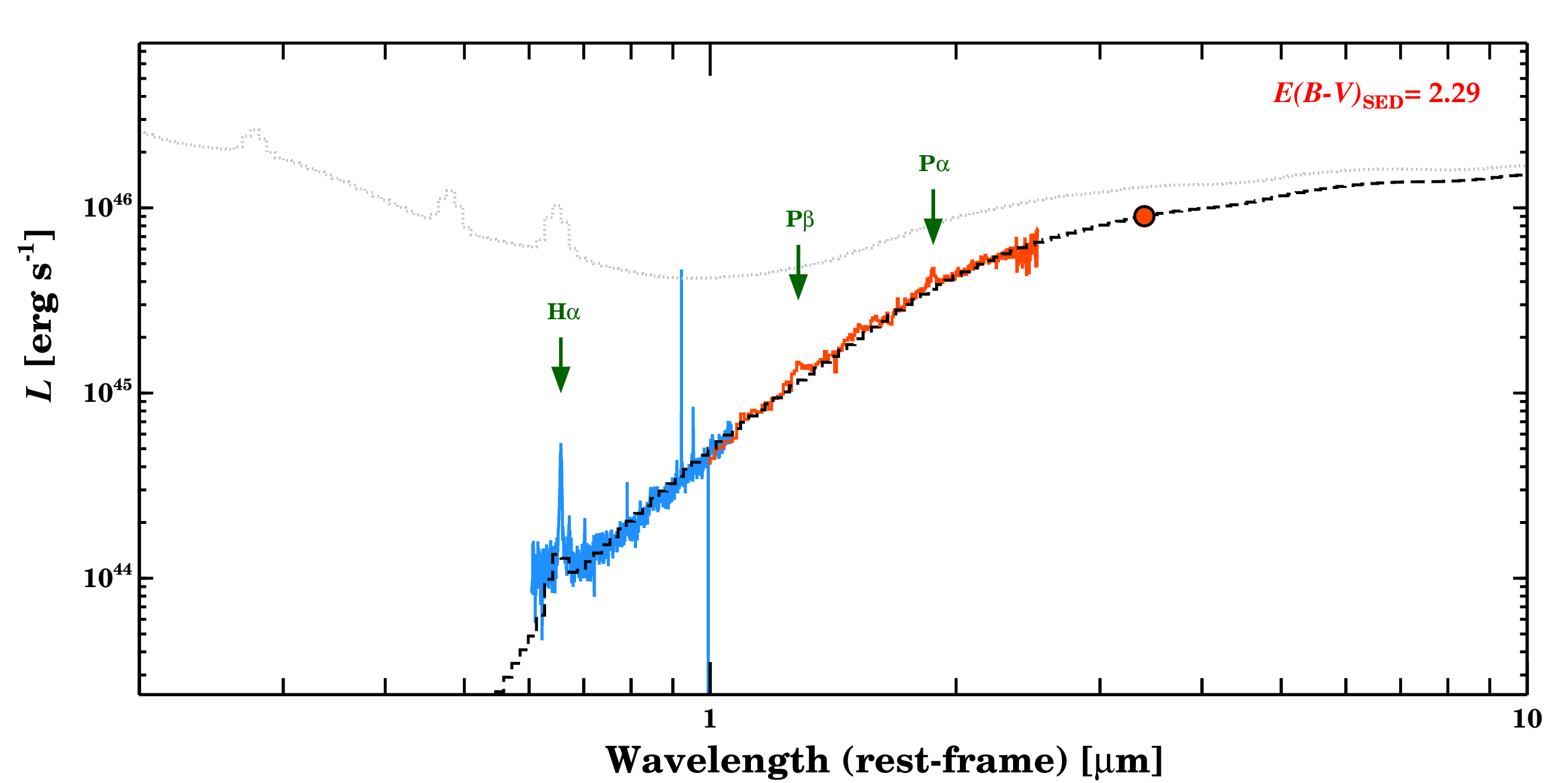}\\
	\caption{
		JWST spectra and best-fit SED model. Blue and red lines represent NIRSepc and MIRI spectra, respectively.
		Black dashed line and gray dotted line indicate the best-fit model and the extinction-corrected best-fit model, respectively.
		Red circle represents $L_{\rm 3.4}$ of the best-fit model.
		H$\alpha$, P$\beta$, and P$\alpha$ lines are marked with green arrows. 
		\label{fig:SED}}
\end{figure*}

For the SED fitting, we use the \texttt{MPFIT} procedure \citep{markwardt09},
and Figure \ref{fig:SED} shows the best-fit model.
Through the SED fitting, we obtain a color excess, $E(B-V)_{\rm SED}$, to be $2.29 \pm 0.29$,
and the extinction-corrected $L_{\rm 3.4}$ is $10^{46.11 \pm 0.06}\,{\rm erg~s^{-1}}$.

\subsection{Emission line fitting} \label{sec:Line}
\begin{figure*}
	\centering
	\figurenum{2}
	\includegraphics[width=\textwidth]{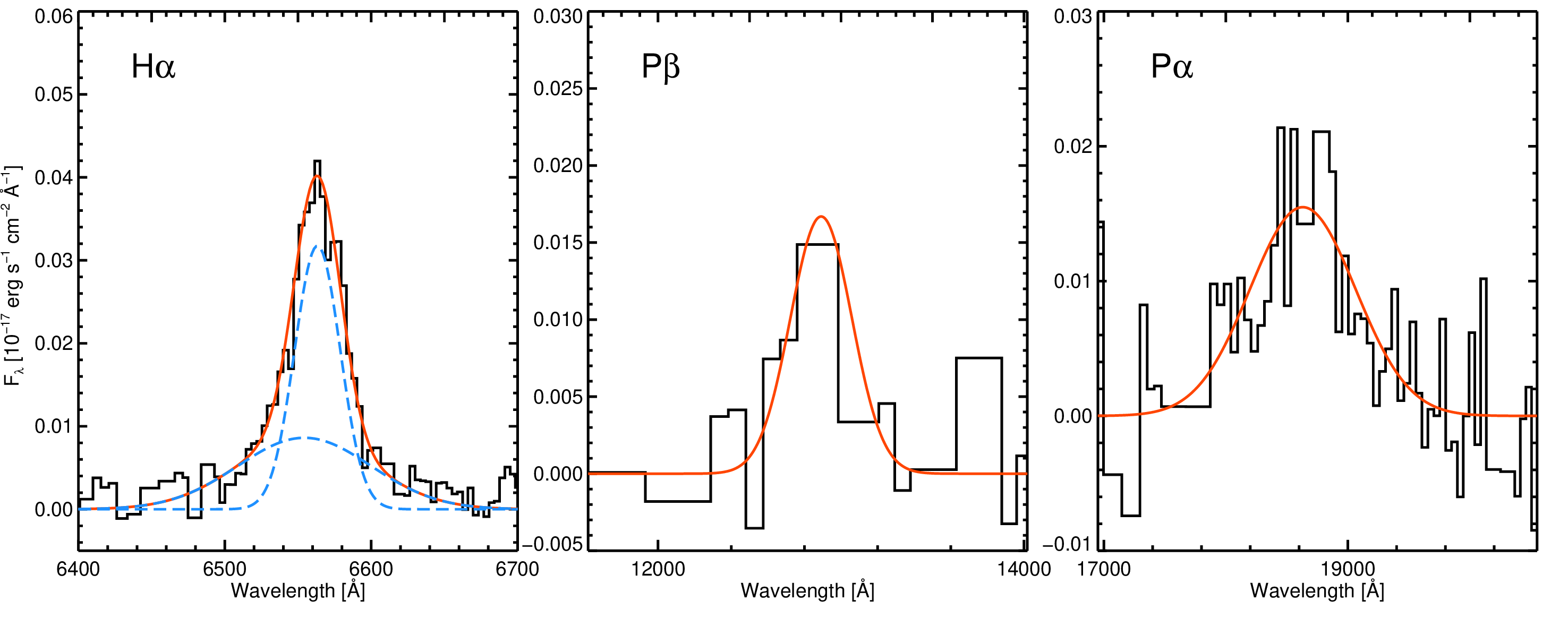}\\
	\caption{
		Fitting results for the H$\alpha$, P$\beta$, and P$\alpha$ lines. 
		The black and red solid lines represent the continuum-subtracted spectra and the best-fit models, respectively.
		The H$\alpha$ line is fit with a double-Gaussian function,
		and each component is presented as the blue dashed line.
		\label{fig:lines}}
\end{figure*}

In this subsection, we describe how H$\alpha$, P$\beta$, and P$\alpha$ lines are fit with the JWST NIRSpec and MIRI spectra.
First, we fit the continua around these lines with a single power law.
After that, the emission lines are fit with a single or double Gaussian function,
and Figure \ref{fig:lines} shows the emission lines and their best-fit models. 

\cite{kim10} showed that the measured fluxes and FWHMs fit with the single, double, and multiple Gaussian functions can be biased,
and provided the correction factors.
The correction factors are $\rm flux_{multi}$/$\rm flux_{single} = 1.06$, $\rm flux_{multi}$/$\rm flux_{double} = 1.05$,
$\rm FWHM_{multi}$/$\rm FWHM_{single} = 0.91$, and $\rm FWHM_{multi}$/$\rm FWHM_{double} = 0.85$.
These factors represent the average ratios, with individual quasars exhibiting values ranging from $\sim$0.75 to $\sim$1.2 \citep{kim10}.
In this study, the measured line properties are corrected to those measured using the multi Gaussian function fitting.

We fit the H$\alpha$ line with a double Gaussian function,
whereas the P$\beta$ and P$\alpha$ lines are fitted with a single Gaussian function
due to the poor S/N and spectral resolution of the spectrum, as shown in Figure \ref{fig:lines}.
The measured FWHM and flux of the H$\alpha$ line are 
1610$\pm$80\,$\rm km~s^{-1}$ and (2.33$\pm$0.20)$\times 10^{-17}\,{\rm erg~s^{-1}~cm^{-2}}$, respectively,
and the measured FWHM value is corrected for the instrumental resolution. 
Note that we also fit the H$\alpha$ line with a single Gaussian function, 
yielding $\frac{\rm FWHM_{H\alpha\,(double)}}{\rm FWHM_{H\alpha\,(single)}}$ and $\frac{\rm flux_{H\alpha\,(double)}}{\rm flux_{H\alpha\,(single)}}$
are $\sim$0.9 and $\sim$1.1, respectively.
These ratios are consistent with the FWHM and flux ratios of the quasars used to derive the correction factors.

For the P$\beta$ and P$\alpha$ lines, we only measure their fluxes due to the poor S/N and spectral resolution of the spectrum.
The measured P$\beta$ and P$\alpha$ line fluxes are 
(7.44$\pm$0.64)$\times 10^{-17}$ and (17.8$\pm$0.9)$\times 10^{-17}\,{\rm erg~s^{-1}~cm^{-2}}$, respectively.
Using these fluxes, we measure the color excess, $E(B-V)_{\rm line}$, to be $2.69 \pm 0.11$ based on the method of \cite{kim18b},
which is a bit larger than the $E(B-V)_{\rm SED}$
but can be taken as a consistent value considering simplified assumptions for the fit.
Note that, in this study, the extinction correction is performed using the $E(B-V)_{\rm SED}$ instead of the $E(B-V)_{\rm line}$
due to the poor quality of the Paschen line spectra. 

\section{Results} \label{sec:result}
\subsection{Bolometric luminosity and BH mass} \label{sec:Lbol&MBH}
To measure the bolometric luminosity, we use the $L_{\rm 3.4}$-based $L_{\rm bol}$ estimator \citep{kim23},
which is
\begin{equation}
	\log \left( \frac{L_{\rm bol}}{\rm 10^{44}\,erg~s^{-1}} \right) = 
	(0.718 \pm 0.011) + \log \left( \frac{L_{\rm 3.4}}{\rm 10^{44}\,erg~s^{-1}} \right). \label{eqn:Lbol_est}
\end{equation}
Using the extinction-corrected $L_{\rm 3.4}$, 
the measured $L_{\rm bol}$ is found to be $10^{46.83\pm0.07}\,{\rm erg~s^{-1}}$.
Additionally, we measure the $L_{\rm bol}$ value using the extinction-corrected $L_{\rm H\alpha}$.
To obtain the $L_{\rm H\alpha}$-based $L_{\rm bol}$, we use several relations between 
the $L_{\rm H\alpha}$, continuum luminosity at 5100\,$\rm \AA{}$ (L5100), and $L_{\rm bol}$.
We bootstrap the relation between $L_{\rm H\alpha}$ and L5100 \citep{greene05},
and between L5100 and $L_{\rm bol}$ \citep{richards06}, as done in \cite{suh24}.
Using the combined relation, the $L_{\rm H\alpha}$-based $L_{\rm bol}$ is found to be $10^{46.46\pm0.04}\,{\rm erg~s^{-1}}$,
which is slightly smaller than the $L_{\rm bol}$ derived from $L_{\rm 3.4}$,
considering the scatter of 0.13\,dex in the relation \citep{kim23}.
Note that, in this study, the $L_{\rm bol}$ from $L_{\rm 3.4}$ is used instead of the $L_{\rm H\alpha}$-based $L_{\rm bol}$.
This is because, even after applying dust-extinction correction, 
the $L_{\rm H\alpha}$-based $L_{\rm bol}$ is sensitive to uncertainties in the $E(B-V)$ measurement,
whereas the $L_{\rm 3.4}$-based $L_{\rm bol}$ is less affected by such uncertainties.

Furthermore, we also measure $L_{\rm bol}$ based on $L_{\rm P\beta}$ and $L_{\rm P\alpha}$
using Paschen-line-based $L_{\rm bol}$ estimators \citep{kim22}.
We use the extinction-corrected Paschen line luminosities,
and the $L_{\rm P\beta}$- and $L_{\rm P\alpha}$-based $L_{\rm bol}$ values are found to be $\sim 10^{46.7}\,{\rm erg~s^{-1}}$,
which are consistent with our $L_{\rm 3.4}$- and $L_{\rm H\alpha}$-based $L_{\rm bol}$ values.

Note that \cite{suh24} also measured the $L_{\rm bol}$ using X-ray data, an SED fitting, and $L_{\rm H\alpha}$.
The $L_{\rm bol}$ values from the X-ray and SED fitting were $\sim 10^{46.6}\,{\rm erg~s^{-1}}$,
which is consistent with our measurements.
However, the $L_{\rm bol}$ from $L_{\rm H\alpha}$ was only $10^{45.6}\,{\rm erg~s^{-1}}$,
which is 10 times smaller than our measurement,
demonstrating the importance of dust-extinction correction 
when deriving physical quantities of quasars from the rest-frame optical spectrum.

We measure $M_{\rm BH}$ based on $L_{\rm 3.4}$ and $\rm FWHM_{H\alpha}$ \citep{kim23}, which is
\begin{equation}
	\begin{aligned}
	\log \left( \frac{M_{\rm BH}}{M_{\odot}} \right) = & (6.952 \pm 0.059) + 
	0.5\log \left( \frac{L_{\rm 3.4}}{\rm 10^{44}\,erg~s^{-1}} \right) \\
	& + (2.06 \pm 0.06) \log \left( \frac{\rm FWHM_{H\alpha}}{\rm 10^3\,km~s^{-1}} \right).
	\end{aligned}
\end{equation}
The measured $M_{\rm BH}$ is $10^{8.43 \pm 0.15}\,M_{\odot}$,
and our measurement is $\sim$40 times higher than 
the $M_{\rm BH}$, $\sim 10^{6.86}\,M_{\odot}$, measured in \cite{suh24}.

Note that \cite{suh24} measured the $M_{\rm BH}$ using the H$\alpha$ line-based estimator \citep{greene05}.
We also measure the $M_{\rm BH}$ using the H$\alpha$ line-based estimator
with the extinction-corrected $L_{\rm H\alpha}$ and the $\rm FWHM_{H\alpha}$.
The obtained $M_{\rm BH}$ is $10^{8.04 \pm 0.19}\,M_{\odot}$,
which is in line with the $L_{\rm 3.4}$-based $M_{\rm BH}$.
\cite{suh24} and we obtained flux and FWHM values of the H$\alpha$ line
that are very similar to each other.
The discrepancy in $M_{\rm BH}$ is mainly caused by differences in extinction correction.

Additionally, we also measure $M_{\rm BH}$ based on $L_{\rm P\beta}$ and $L_{\rm P\alpha}$
using the Paschen-line-based estimators \citep{kim22}.
To obtain the Paschen-line-based $M_{\rm BH}$, Paschen line luminosity and FWHM values are necessary,
and the $\rm FWHM_{P\beta}$ and $\rm FWHM_{P\alpha}$ values are converted from the $\rm FWHM_{H\alpha}$ \citep{kim10}.
With the extinction-corrected $L_{\rm P\beta}$ and $L_{\rm P\alpha}$,
we measure the P$\beta$- and P$\alpha$-based $M_{\rm BH}$ values, which are $\sim 10^{8.2}\,M_{\odot}$.
These measurements agree with the $L_{\rm 3.4}$- and $L_{\rm H\alpha}$-based $M_{\rm BH}$ values,
and significantly higher than that of \cite{suh24}.

\subsection{Eddington ratio} \label{sec:REdd}
We measure $\lambda_{\rm Edd}$ using the newly measured $L_{\rm bol}$ and $M_{\rm BH}$ values.
If we use the $L_{\rm 3.4}$-based $M_{\rm BH}$,
$\lambda_{\rm Edd}$ is found to be 1.97$\pm$0.88,
suggesting that LID-568 exhibits super-Eddington BH accretion activity only at a modest level.
However, our measurement is significantly smaller than the $\lambda_{\rm Edd}$ of 41.5 measured in \cite{suh24}.
This discrepancy is mainly due to the differences in the $M_{\rm BH}$ measurements. 
Note that, if the $L_{\rm H\alpha}$-based $M_{\rm BH}$ is adopted,
$\lambda_{\rm Edd}$ increases to 4.83$\pm$2.86, similar to
the H$\alpha$-based $\lambda_{\rm Edd}$ value in \cite{suh24}.

\begin{figure*}
	\centering
	\figurenum{3}
	\includegraphics[width=\textwidth]{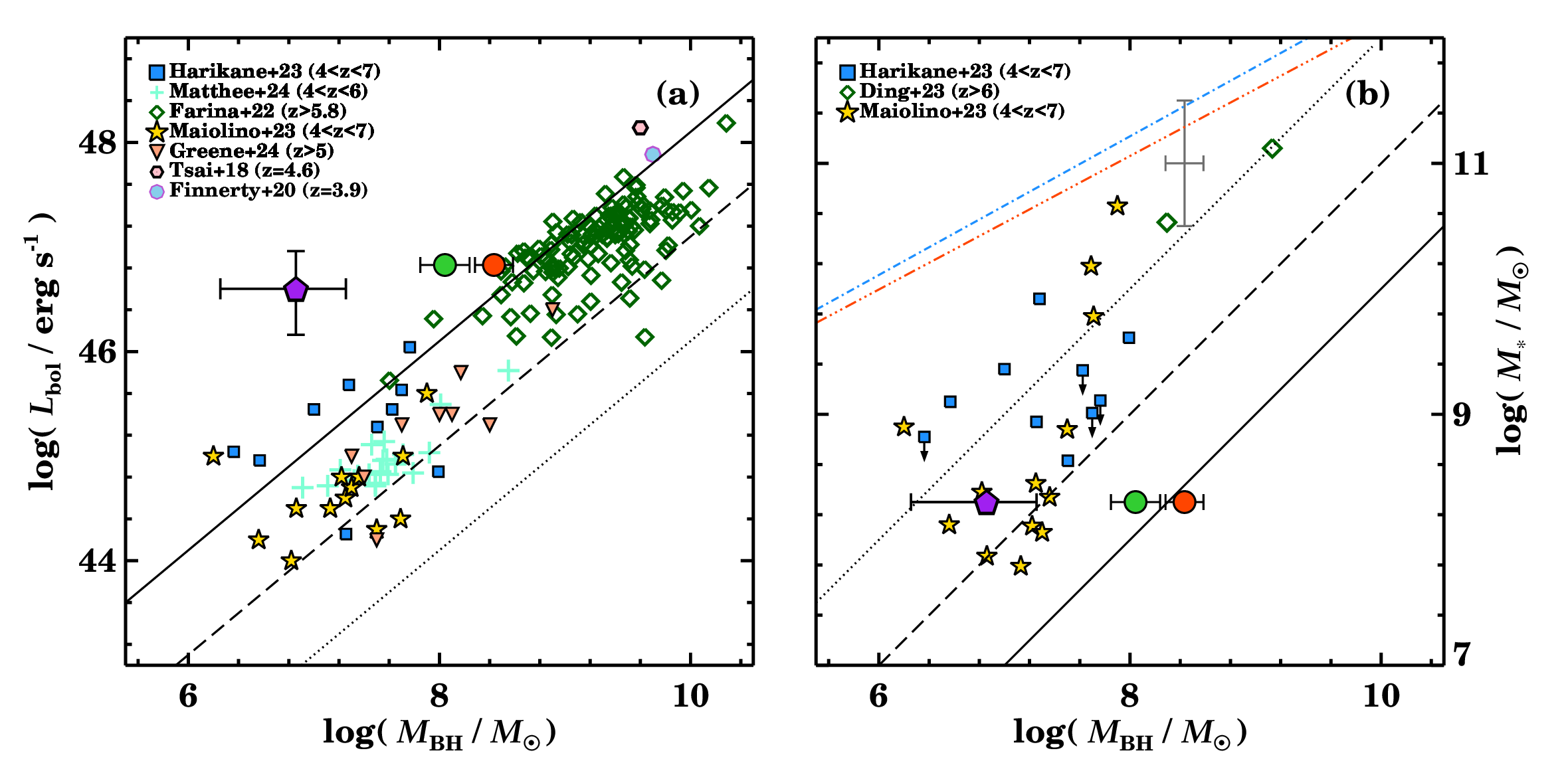}\\

	\caption{
		(a) $M_{\rm BH}$ vs. $L_{\rm bol}$ of LID-568 and quasars at similar redshifts.
		LID-568 is marked with circles, where the red and green circles represent
		the $L_{\rm 3.4}$-based $M_{\rm BH}$ and the $L_{\rm H\alpha}$-based $M_{\rm BH}$, respectively.
		The purple pentagon indicates the values measured in \cite{suh24}.
		The solid, dashed, and dotted lines indicate $\lambda_{\rm Edd}$ of 1, 0.1, and 0.01, respectively.
		(b) Comparison of $M_{\rm BH}$ and $M_{\ast}$. 
		The meanings of the red and green circles and purple pentagon are identical to those in panel (a).
		Gray error bar represents the $L_{\rm 3.4}$-based $M_{\rm BH}$
		and the $M_{\ast}$ from the SFR of the host galaxy.
		The solid, dashed, and dotted lines mean $M_{\rm BH}$-to-$M_{\ast}$ ratio of 1, 0.1, and 0.01, respectively,
		and the dot-dashed blue and dot-dot-dot-dashed red lines represent 
		disk-dominated and bulge-dominated AGNs in the nearby Universe \citep{zhuang23}.		
		\label{fig:prop}}
\end{figure*}

In Figure \ref{fig:prop}a, we compare the $L_{\rm bol}$ versus $M_{\rm BH}$ of LID-568 with quasars at $z \sim$4--7.
The measured $\lambda_{\rm Edd}$ of LID-568 is comparable to
the highest $\lambda_{\rm Edd}$ values of quasars at $z \sim$4--7 (\citealt{tsai18,finnerty20,harikane23,matthee24,farina22,maiolino23,greene24}).
Note that these comparisons cover various types of quasars, including dust-obscured quasars.
Overall, our result indicates that LID-568 harbors a rapidly growing SMBH at around the Eddington limit.

Finally, we remark that recent observational studies (e.g., \citealt{du16,du19,gravity24}) have shown that
quasars with high luminosities and high $\lambda_{\rm Edd}$ values could have smaller broad line region (BLR) sizes by factors of a few,
compared to those expected from well-known $R_{\rm BLR}$--$L$ relationships.
If this is true, high-$\lambda_{\rm Edd}$ quasars tend to have overestimated $M_{\rm BH}$ values,
implying that LID-568's $M_{\rm BH}$ and $\lambda_{\rm Edd}$ may deviate from the measured quantities,
tending toward a lower $M_{\rm BH}$ and higher $\lambda_{\rm Edd}$.
However, in such a case, all other points near $\lambda_{\rm Edd} \sim 1$ would move leftward 
(higher $\lambda_{\rm Edd}$ values by a factor of a few) in Figure \ref{fig:prop}.

\subsection{Host galaxy property} \label{sec:host}
It is noteworthy that our $M_{\rm BH}$ value is comparable to 
the stellar mass ($M_{\ast}$) of the host galaxy \citep{suh24}.
\cite{suh24} performed far-infrared (FIR) SED fitting extending to hundreds of microns,
using a power-law and two greybody components \citep{casey12},
and measured a total IR luminosity ($L_{\rm 8-1,000\,\mu m}$) and dust mass of $\sim 2.95\times 10^{6}\,M_{\odot}$, respectively.
Using the dust-to-stellar mass ratio at a similar redshift \citep{xiao23}, the $M_{\ast}$ is found to be $\sim 2\times 10^{8}\,M_{\odot}$.

Alternatively, we estimated $M_{\ast}$ based on star-formation rate (SFR) of the host galaxy. 
According to \cite{suh24}, the observed specific flux at 870\,$\mu$m of LID-568 is 545$\pm$158\,$\mu$Jy. 
Assuming that the rest-frame FIR flux is dominated by star formation, 
and approximating that the rest-frame wavelength of 175\,$\mu$m of the sub-mm band to 160\,$\mu$m, 
we obtain SFR of $\sim$ 1000\,$M_{\odot}~{\rm yr^{-1}}$ adopting the FIR-SFR correlation of \cite{calzetti10}. 
From the observed SFR vs. $M_{\ast}$ of starburst or luminous infrared galaxies \citep{rodighiero11}, 
we get $\log \left( M_{\ast} / M_{\odot} \right) =$10.5 to 11.5.

Based on $M_{\ast}$ from \cite{suh24},
LID-568 has an $M_{\rm BH}$-to-$M_{\ast}$ ratio of $\sim$1, 
significantly higher than those of other quasars at similar redshifts \citep{maiolino23,ding23,harikane23} or 
AGNs in the nearby Universe \citep{zhuang23}, as shown in Figure \ref{fig:prop}b.
On the other hand, if we adopt our estimate of $M_{\ast}$, 
the $M_{\rm BH}$ to $M_{\ast}$ ratio of LID-568 is comparable to that of 
other quasars at similar redshifts. 
More reliable measurements of $M_{\ast}$ should resolve the differences in the $M_{\rm BH}$ to $M_{\ast}$ ratios.

\section{Summary} \label{sec:summary}
We reanalyzed the JWST spectra of LID-568, an intriguing dust-obscured quasar at $z \sim 4$
whose SMBH has been suggested to be undergoing a super-Eddington accretion at $\lambda_{\rm Edd} \sim 40$. 
However, our analysis that accounts for the heavy dust-extinction reveals that LID-568
seems undergoing Eddington-limited accretion at $\lambda_{\rm Edd} \sim 1$. 
Our new $M_{\rm BH}$ value is about $10^{8.43}$\,$M_{\odot}$, $\sim$40 times higher than that suggested by \cite{suh24},
which is the reason for the discrepancy.
These results emphasize the challenges and importance of dust-extinction correction
in understanding the nature of dust-obscured quasars.
Moreover, our IR-based analysis could better constrain the nature of newly discovered high-$z$ dust-obscured quasars --
often referred to as ``little red dots" and found through JWST observations.

\begin{acknowledgments}
We thank Hyewon Suh for useful discussions, as well as the anonymous referee for constructive comments.
This work is supported by the National Research Foundation of Korea (NRF) grant, 
No. 2020R1A2C3011091 and 2021M3F7A1084525, funded by the Korea government (MSIT).
D.K. acknowledges the support by the National Research Foundation of Korea (NRF) grant 
(No. 2021R1C1C1013580 and 2022R1A4A3031306) funded by the Korean government (MSIT).
\end{acknowledgments}

\clearpage

\end{document}